\documentclass{article}
\usepackage{nips07submit_e,times}
\usepackage{amsmath}
\usepackage{amsfonts}
\usepackage{amssymb}
\usepackage[dvips]{graphicx}

\title{Self-organization using synaptic plasticity}

\author{
Vicen\c{c} G\'omez${}^{1}$\\
\texttt{vgomez@iua.upf.edu} \\
\And
Andreas Kaltenbrunner${}^{2}$\\
\texttt{andreas.kaltenbrunner@upf.edu} \\
\And
Hilbert J Kappen${}^{1}$\\
\texttt{b.kappen@science.ru.nl} \\
\And	
      Vicente L\'opez${}^{2}$\\
\texttt{vicente.lopez@barcelonamedia.org} \\
\And
${}^{1}$Department of Biophysics\\
Radboud University Nijmegen\\
6525 EZ Nijmegen, The Netherlands\\
\And
${}^{2}$Barcelona Media -– Innovation Centre \\
Av. Diagonal 177,\\
08018 Barcelona, Spain\\
}

\begin{document}

%\makeanontitle
\maketitle

\begin{abstract}
Large networks of spiking neurons show abrupt changes in their collective dynamics
resembling phase transitions studied in statistical physics. An example of this
phenomenon is the transition from irregular, noise-driven dynamics to regular,
self-sustained behavior observed in networks of integrate-and-fire
neurons as the interaction strength between the neurons increases. 
In this work we show how a network of spiking neurons
is able to self-organize towards a critical state for which
the range of possible inter-spike-intervals (dynamic range) is maximized.
Self-organization occurs via synaptic dynamics that we analytically derive.
The resulting plasticity rule is defined locally so that global homeostasis near the critical
state is achieved by local regulation of individual synapses.
\end{abstract}
\section{Introduction}
It is accepted that neural activity self-regulates to prevent neural circuits
from becoming hyper- or hypoactive by means of homeostatic processes
\cite{Turrigiano}.  Closely related to this idea is the claim that optimal
information processing in complex systems is attained at a critical point, near
a transition between an ordered and an unordered regime of dynamics
\cite{Bertschinger,Packard, Langton}.  Recently, Kinouchi and %Legenstein07,
Copelli~\cite{Kinouch2006} provided  a realization of this claim, showing that
sensitivity and dynamic range of a network are maximized at the critical point
of a non-equilibrium phase transition.  Their findings may explain how
sensitivity over high dynamic ranges is achieved by living organisms.

Self-Organized Criticality (SOC) \cite{Bak} has been proposed as a mechanism
for neural systems which evolve \emph{naturally} to a critical state without
any tuning of external parameters. In a critical state, typical macroscopic
quantities present structural or temporal scale-invariance.  Experimental
results~\cite{Beggs03} show the presence of neuronal avalanches of scale-free
distributed sizes and durations, thus giving evidence of SOC under suitable
conditions.  A possible regulation mechanism may be provided by synaptic
plasticity, as proposed in \cite{Levina07}, where synaptic depression is
shown to cause the mean synaptic strengths to approach a critical value for a
range of interaction parameters which grows with the system size.

In this work we analytically derive a local synaptic rule that can drive and
maintain a neural network near the critical state. According to the proposed
rule, synapses are either strengthened or weakened whenever a post-synaptic
neuron receives either more or less input from the population than the required
to fire at its \emph{natural} frequency. This simple principle is enough for
the network to self-organize at a critical region where the dynamic range is
maximized.  We illustrate this using a model of non-leaky spiking neurons with
delayed coupling for which a phase transition was analyzed
in~\cite{Kaltenbrunner2007NECO}.
\section{The model}
The model under consideration was introduced in \cite{RODR01} and can be
considered as an extension of~\cite{VanvreeswijkA93,GERSTEIN64}.
The state of a neuron $i$ at time $t$ is encoded by its activation
level $a_i(t)$, which performs at discrete timesteps a random walk
with positive drift towards an absorbing barrier $L$. This
\emph{spontaneous} evolution is modelled using a Bernoulli process
with parameter $p$.  When the threshold $L$ is reached, the states of
the other units $j$ in the network are increased after one timestep
by the synaptic efficacy $\epsilon_{ji}$, $a_i$ is reset to $1$, and
the unit $i$ remains insensitive to incoming spikes during the
following timestep.
The evolution of a neuron $i$ can be described by the following recursive rules:
\begin{align}\label{eq:discrete-final}
a_i(t+1) & =
\begin{cases}
  a_i(t) + \displaystyle \sum_{j=1,j\neq i}^N \epsilon_{ij} H_L(a_j(t)) + 1 & \text{with probability $p$}\\
  a_i(t) + \displaystyle \sum_{j=1,j\neq i}^N \displaystyle \epsilon_{ij} H_L(a_j(t))     & \text{with probability $1-p$}\\
\end{cases} & \text{if $a_i(t) < L$}\notag\\
a_i(t+1) & = 1 + \sum_{j=1,j\neq i}^N \epsilon_{ij} H_L(a_j(t)) & \text{if
  $a_i(t) \geq L$}
\end{align}
where $H_L(x)$ is the Heaviside step function: $H_L(x) = 1$ if $x\geq L$, and
$0$ otherwise.

\begin{figure}[!t]
\begin{center}
  \includegraphics[width=.45\columnwidth,angle=-90]{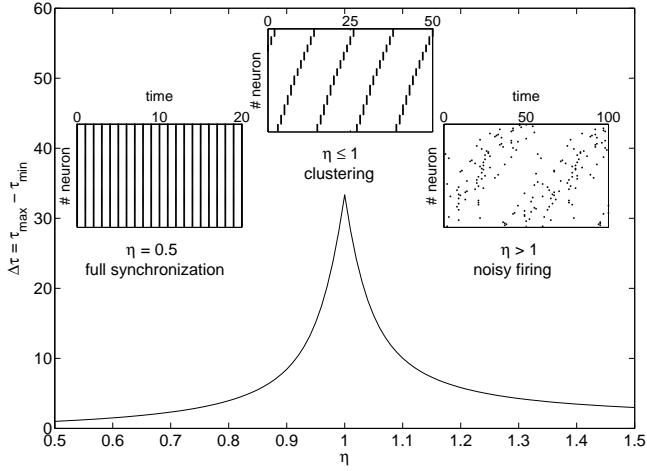}
\end{center}
\caption[Dynamics of the model]
{
Number of possible ISIs according to the bound $\Delta\tau =
	\tau_{max}-\tau_{min}$ derived in~\cite{Kaltenbrunner2007NECO}.  
	For~$\eta>1$
	the network presents sub-critical behavior and is dominated by the noise. For
	$\eta<1$~it shows super-critical behavior. Criticality is produced at
	$\eta=1$, which coincides to the onset of sustained activity. At this point, the network is also broken down in a maximal
	number of clusters of units which fire according to a periodic
	pattern.
\label{fig:period-range}
}
\end{figure}

Using the mean synaptic efficacy: $\langle\epsilon\rangle=\sum_i^N\sum_{j,j\neq
i}^N\epsilon_{ij}/(N(N-1))$ we describe the degree of interaction between the
units with the following characteristic parameter:
\begin{align} \label{eq:eta}
\eta = \frac{L-1}{(N-1)\langle\epsilon\rangle},
\end{align}
which indicates whether the spontaneous dynamics ($\eta > 1$) or the message
interchange mechanism ($\eta \leq 1$) dominates the behavior of the system. As
illustrated in the right raster-plot of Figure~\ref{fig:period-range}, at
$\eta> 1$ neurons fire irregularly as independent oscillators, whereas at
$\eta=1$ (central raster-plot) they synchronize into several phase-locked
clusters.  The lower $\eta$, the less clusters can be observed. For $\eta=0.5$
the network is fully synchronized (left raster-plot).

In \cite{Kaltenbrunner2007NECO} it is shown that the system undergoes a phase
transition around the critical value $\eta=1$.  The study provides upper
($\tau_{max}$) and lower bounds ($\tau_{min}$) for the mean
inter-spike-interval (ISI) $\tau$ of the ensemble and shows that the range of
possible ISIs taking the average network behavior
($\Delta\tau=\tau_{max}$-$\tau_{min}$) is maximized at $\eta=1$. This is
illustrated in Figure~\ref{fig:period-range} and has been observed as well
in~\cite{Kinouch2006} for a similar neural model.

The average of the mean ISI $\langle\tau\rangle$ is of order $N^x$ with
exponent $x=1$ for $\eta > 1$, $x=0.5$ for $\eta =1$, and $x=0$ for $\eta < 1$
as $N\rightarrow\infty$, and can be approximated as shown in
~\cite{Kaltenbrunner2007NECO} with \footnote{The equation was denoted
$\langle\tau\rangle_{min}$ in \cite{Kaltenbrunner2007NECO}. We slightly
modified it using $\langle\epsilon\rangle$ and replacing $\eta$ by
Eq.~\eqref{eq:eta}.}:
%\emph{Compare here the model of Geisel and Copelli}.
%Apart form the bounds
%and can be approximated with \cite{Kaltenbrunner2007NECO}.
%$\langle\tau\rangle$ was given as well in
%We will use this formula for the subsequent analysis.
% \begin{align}\label{con:tau_approx} 
%   \langle\tau\rangle =
%   1+\frac{(N-1)\langle\epsilon\rangle(\eta-1)-\langle\epsilon\rangle}{2p}
%   +\sqrt{\left(\frac{(N-1)\langle\epsilon\rangle(\eta-1) -\langle\epsilon\rangle}{2p}
%       +1\right)^2+\frac{N\langle\epsilon\rangle}{2p}}
% \end{align}
% \begin{align}\label{con:tau_approx}
%   \langle\tau\rangle_{min}  = 1+
%   \frac{L-1-(N-1)\langle\epsilon\rangle-\langle\epsilon\rangle}{2p}+\sqrt{ \left(
%       \displaystyle \frac{L-1-(N-1)\langle\epsilon\rangle-\langle\epsilon\rangle}{2p}
%       + 1 \right)^2 + \frac{N\langle\epsilon\rangle}{2p} }.
% \end{align}
\begin{align}\label{con:tau_approx}
  \tau_{app}  = 1+
  \frac{L-1-N\langle\epsilon\rangle}{2p}+\sqrt{ \left(
      \displaystyle \frac{L-1-N\langle\epsilon\rangle}{2p}
      + 1 \right)^2 + \frac{N\langle\epsilon\rangle}{2p} }.
\end{align}

\section{Self-organization using synaptic plasticity}
\label{sec:learning}
We now introduce synaptic dynamics in the model.  We first present the
\emph{dissipated spontaneous evolution}, a magnitude also maximized at
$\eta=1$.  The gradient of this magnitude turns to be simple analytically and
leads to a plasticity rule that can be expressed using only local information
encoded in the post-synaptic unit.
\subsection{The dissipated spontaneous evolution}
During one ISI, we distinguish between the spontaneous evolution carried out by
a unit and the actual spontaneous evolution needed for a unit to reach the
threshold $L$.  The difference of both quantities can be regarded as a surplus
of spontaneous evolution, which is dissipated during an ISI.

\begin{figure}[!t]
\begin{center}
\includegraphics[width=.5\textwidth,angle=-90]{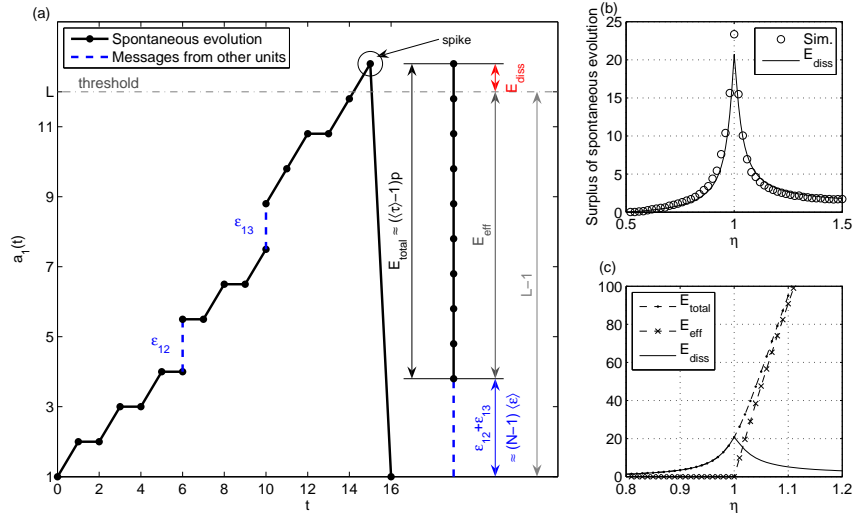}
\end{center}
{\caption[Dissipated spontaneous evolution $E_{diss}$]
{\textbf{(a)} Example trajectory of the state of a neuron: the dissipated spontaneous evolution $E_{diss}$ is the difference
between the total spontaneous evolution $E_{total}$ (thick black lines)
and the actual evolution required to reach the threshold $E_{eff}$ (dark gray dimensioning) in one ISI.
\textbf{(b)} $E_{diss}$ is maximized at the critical point.
\textbf{(c)}  The three different evolutions involved in the analysis
  (parameters for (b) and (c) are $N=L=1000$ and $p=0.9$.  For the mean ISI we used $\tau_{app}$
  of Eq.~\eqref{con:tau_approx}).
  %\tau \rangle_{min}$ derived in \cite{Kaltenbrunner2007NECO}.
  % the previous chapter.  }
}
%
%  $N=L=1000$ and $p=0.9$.  For the mean period we used $\langle
\label{fig:remaining}}
\end{figure}

Figure \ref{fig:remaining}a shows an example trajectory of a neuron's state.
First, we calculate the spontaneous evolution of the given unit during one ISI,
which it is just its number of stochastic state transitions during an ISI of
length~$\tau$ (thick black lines in Figure \ref{fig:remaining}a).  These state
transitions occur with probability $p$ at every timestep except from the
timestep directly after spiking.  Using the average ISI-length
$\langle\tau\rangle$ over many spikes and all units we can calculate the
average total spontaneous evolution:
\begin{align}\label{eq:e_total} 
%E_{total} & = (\langle\tau\rangle-t_{ref})p.
E_{total} & = (\langle\tau\rangle-1)p.
\end{align} 
Since the state of a given unit can exceed the threshold because of the
received messages from the rest of the population (blue dashed lines in Figure
\ref{fig:remaining}a), a fraction of \eqref{eq:e_total} is actually not
required to induce a spike in that unit, and therefore is dissipated.  We can
obtain this fraction by subtracting from \eqref{eq:e_total} the actual number
of state transitions that was required to reach the threshold $L$.  The latter
quantity can be referred to as \emph{effective} spontaneous evolution $E_{eff}$
and is on average $L-1$ minus $(N-1)\langle\epsilon\rangle$, the mean evolution
caused by the messages received from the rest of the units during an ISI.  For
$\eta \leq 1$, the activity is self-sustained and the messages from other units
are enough to drive a unit above the threshold. In this case, all the
spontaneous evolution is dissipated and $E_{eff}=0$.  Summarizing, we have
that:
\begin{align}\label{eq:e_eff1}
E_{eff} & = \max\{0, L-1-(N-1)\langle\epsilon\rangle\}=\begin{cases}
              L-1-(N-1)\langle\epsilon\rangle & \text{for $\eta \geq 1$} \\
              0                 & \text{for $\eta < 1$}
            \end{cases}
\end{align}
If we subtract \eqref{eq:e_eff1} from $E_{total}$ \eqref{eq:e_total}, we obtain
the mean dissipated spontaneous evolution, which is visualized as red
dimensioning in Figure \ref{fig:remaining}a:
\begin{align}\label{eq:free}
E_{diss} & = E_{total} - E_{eff} = (\langle\tau\rangle-1)p
-\max\{0, L-1-(N-1)\langle\epsilon\rangle\}.
\end{align}
Using~\eqref{con:tau_approx} as an approximation of $\langle \tau \rangle$ we
can get an analytic expression for $E_{diss}$.
%\begin{multline}\label{con:tau_limit_min} 
% {\langle \tau \rangle}_{min} = \frac{(N-1)\epsilon(\eta-1)
%-\epsilon}{2p} +\frac{1+t_{ref}}{2} +\\
% \sqrt{\left(\frac{(N-1)\epsilon(\eta-1)
%-\epsilon}{2p} +\frac{1+t_{ref}}{2}\right)^2+\frac{N\epsilon\delta}{p \langle g \rangle}}
%\end{multline}
%Then:
%For clarity and without loss of generality, we assume the same values for
%transmission delay and the refractory period in all the units, $t_{ref} =
%\delta = 1$. Then $E_{diss}$ in our model is:
%
% \begin{multline}\label{con:tau_limit_min2} 
% E_{diss} =  \left(\frac{(N-1)\epsilon(\eta-1)-\epsilon}{2p} + \sqrt{\left(\frac{(N-1)\epsilon(\eta-1)
%  -\epsilon}{2p} +1\right)^2+\frac{N\epsilon}{2p}}\right)p \\
%  -\max\{0, L-1-(N-1)\epsilon\}.
% \end{multline}
% \begin{align}\label{con:tau_limit_min2} 
% E_{diss} =  \left(\frac{(N-1)\langle\epsilon\rangle(\eta-1)-\langle\epsilon\rangle}{2p} + \sqrt{\left(\frac{(N-1)\langle\epsilon\rangle(\eta-1)
%  -\langle\epsilon\rangle}{2p} +1\right)^2+\frac{N\langle\epsilon\rangle}{2p}}\right)p - E_{eff}.
% \end{align}
Figures~\ref{fig:remaining}b and c show this analytic curve $E_{diss}$ in
function of $\eta$ together with the outcome of simulations.

At $\eta>1$ the units reach the threshold $L$ mainly because of their
spontaneous evolution. Hence, $E_{total}\approx E_{eff}$ and $E_{diss} \approx
0$.  The difference between $E_{total}$ and $E_{eff}$ increases as $\eta$
approaches~$1$ because the message interchange progressively dominates the
dynamics.  At $\eta<1$, we have $E_{eff}=0$.  In this scenario $E_{diss} =
E_{total}$, is mainly determined by the ISI $\langle\tau\rangle$ and thus
decays again for decreasing $\eta$.  The maximum can be found at $\eta=1$.
\subsection{Synaptic dynamics}
After having presented our magnitude of interest we now derive a plasticity
rule in the model.  Our approach essentially assumes that updates of the
individual synapses $\epsilon_{ij}$ are made in the direction of the gradient
of $E_{diss}$.  The analytical results are rather simple and allow a clear
interpretation of the underlying mechanism governing the dynamics of the
network under the proposed synaptic rule.

We start approximating the terms
$N\langle\epsilon\rangle$ and $(N-1)\langle\epsilon\rangle$ by the sum
of all pre-synaptic efficacies $\epsilon_{ik}$:
\begin{align}\label{eq:suma-eps1} 
  N\langle\epsilon\rangle=(N-1)\langle\epsilon\rangle +\langle\epsilon\rangle & \approx
  (N-1)\langle\epsilon\rangle = \sum_{i=1}^N\sum_{k\neq i}{\epsilon_{ik}}/N
  \approx \sum_{k\neq i}{\epsilon_{ik}}.
\end{align}
This can be done for large $N$ and if we suppose that the distribution of
$\epsilon_{ik}$ is the same for all $i$. $E_{diss}$ is now defined in terms of
each individual neuron $i$ as:
\begin{multline}\label{eq:e_free_eps2}
E_{diss}^{i}  = \left( \frac{L-1-\sum_{k\neq i}{\epsilon_{ik}}}{2p}+\sqrt{ \left( \displaystyle
\frac{L-1-\sum_{k\neq i}{\epsilon_{ik}}}{2p} + 1 \right)^2 + \frac{\sum_{k\neq i}{\epsilon_{ik}}}{2p} } \right)p\\
- \max\{0, L-1-\sum_{k\neq i}{\epsilon_{ik}}\}.
\end{multline}
%AK: this is not clear to me. What would other schedulings be?
An update of $\epsilon_{ij}$ occurs when a spike from the pre-synaptic unit $j$
induces a spike in a post-synaptic unit $i$.  Other schedulings are also
possible. The results are robust as long as synaptic updates are produced at
the spike-time of the post-synaptic neuron.
\begin{align}\label{eq:rule}
\Delta \epsilon_{ij} = \kappa \frac{\partial{E^i_{diss}}}{\partial{\epsilon_{ij}}}= \kappa \left( \frac{\partial{E^i_{total}}}{\partial{\epsilon_{ij}}} - \frac{\partial E^i_{eff}}{\partial{\epsilon_{ij}}}\right),
\end{align}
where the constant $\kappa$ scales the amount of change in the synapse.
%In the next section we will analyze the role of a global $\kappa$ in the system.
% For the moment, let us calculate the derivative $E^i_{diss}$ with respect to
% an arbitrary synapse $\epsilon_{ij}$.  On one hand, differentiating
% $E^i_{total}$ yields to:
% \begin{align}\label{eq:derivation}
% \frac{\partial{E^i_{total}}}{\partial{\epsilon_{ij}}}
% %& = \left( -\frac{\delta_{kj}}{2p} + \frac{2\left( \frac{L-1-\sum_{k\neq i}{\epsilon_{ik}}}{2p}+1 \right)\left(-\frac{\delta_{kj}}{2p}\right) +\frac{\delta_{kj}}{2p} }{2\sqrt{ \left(
% %\frac{L-1-\sum_{k\neq i}{\epsilon_{ik}}}{2p} + 1 \right)^2 + \frac{\sum_{k\neq i}{\epsilon_{ik}}}{2p} } }\right) p \\
% %& = \frac{\delta_{kj}}{2}\left( -1 + \frac{-2\left( \frac{L-1-\sum_{k\neq i}{\epsilon_{ik}}}{2p} +1\right)+1  }{2\sqrt{ \left(
% %\frac{L-1-\sum_{k\neq i}{\epsilon_{ik}}}{2p} + 1 \right)^2 + \frac{\sum_{k\neq i}{\epsilon_{ik}}}{2p} } }  \right)\\
% & = -\frac{1}{2}\left( \frac{ \frac{L-1-\sum_{k\neq i}{\epsilon_{ik}}}{2p}+\frac{1}{2} }{ \sqrt{ \left(
% \frac{L-1-\sum_{k\neq i}{\epsilon_{ik}}}{2p} + 1 \right)^2 + \frac{\sum_{k\neq i}{\epsilon_{ik}}}{2p} }} + 1\right),
% \end{align}
% On the other hand, the derivative of ${E^i_{eff}}$ with respect to $\epsilon_{ij}$ is:
% \begin{align*}
% \frac{\partial E^i_{eff}}{\partial{\epsilon_{ij}}} = \begin{cases}
%     0 & \text{if $(L-1 - \sum_{k\neq i}{\epsilon_{ik}}) < 0$} \\
%     \text{indef} & \text{if $(L-1 - \sum_{k\neq i}{\epsilon_{ik}}) = 0$} \\
%     -1 & \text{if $(L-1 - \sum_{k\neq i}{\epsilon_{ik}}) > 0$}.
%   \end{cases}
% \end{align*}
%We can write the gradient thus as:
We can write the gradient as:
\begin{align}\label{eq:derivation}
   \frac{\partial{E^i_{diss}}}{\partial{\epsilon_{ij}}} = \frac{ -\frac{1}{2}
    \left(\frac{L-1-\sum_{k\neq i}{\epsilon_{ik}}}{2p}+\frac{1}{2}
    \right)}{ \sqrt{ \left( \frac{L-1-\sum_{k\neq
            i}{\epsilon_{ik}}}{2p} + 1 \right)^2 + \frac{\sum_{k\neq
          i}{\epsilon_{ik}}}{2p} }} -\frac{1}{2} -
  \begin{cases}
    0 & \text{if $(L-1 - \sum_{k\neq i}{\epsilon_{ik}}) < 0$} \\
    \text{indef} & \text{if $(L-1 - \sum_{k\neq i}{\epsilon_{ik}}) = 0$} \\
    -1 & \text{if $(L-1 - \sum_{k\neq i}{\epsilon_{ik}}) > 0$}.
  \end{cases}
\end{align}
For a plasticity rule to be biologically plausible it must be local, so only
information encoded in the states of the pre-synaptic $j$ and the post-synaptic
$i$ neurons must be considered to update $\epsilon_{ij}$.

We propagate $\sum_{k\neq i}{\epsilon_{ik}}$ to the state of the post-synaptic
unit $i$ by considering for every unit $i$, an effective threshold $L^i$ which
decreases deterministically every time an incoming pulse is
received~\cite{gomez2006}.  At the end of an ISI $L^i\approx(L-1 - \sum_{k\neq
i}{\epsilon_{ik}})$ and encodes implicitly all pre-synaptic efficacies of~$i$.
Intuitively, $L^i$ indicates how the activity received from the population in
the last ISI differs from the activity required to induce and spike in $i$.

The only term involving non-local information in~\eqref{eq:derivation} is the
noise rate $p$. We replace it by a constant $c$ and show later its limited
influence on the synaptic rule.  With these modifications we can write the
derivative of $E^i_{diss}$ with respect to $\epsilon_{ij}$ as a function of
only local terms:
% \begin{align}\label{eq:gradient}
% \frac{\partial{E^i_{diss}}}{\partial{\epsilon_{ij}}}
% & =  \frac{ -\frac{L^i}{2c}-\frac{1}{4} }{ \sqrt{ \left(
% \frac{L^i}{c} + 1 \right)^2 + \frac{L-L^i}{c} }} - \frac{1}{2}+ \begin{cases}
%     0 & \text{if $L^i < 0$} \\
%     \text{indef} & \text{if $L^i = 0$} \\
%     1 & \text{if $L^i > 0$}.
%   \end{cases}
% \end{align}
\begin{align}\label{eq:gradient}
\frac{\partial{E^i_{diss}}}{\partial{\epsilon_{ij}}}
& =  \frac{ -L^i-c }{ 2\sqrt{ \left(
L^i + 2c \right)^2 + 2c(L-L^i) }} + \frac{\text{sgn}(L^i)}{2}
\end{align}
Note that, although the derivation based on the surplus spontaneous evolution
\eqref{eq:derivation} may involve information not locally accessible to the
neuron, the derived rule \eqref{eq:gradient} only requires a mechanism to keep
track of the difference between the natural ISI and the actual one.

We can understand the mechanism involved in a particular synaptic update by
analyzing in detail Eq.~\eqref{eq:gradient}.  In the case of a negative
effective threshold ($L^i < 0$) unit $i$ receives more input from the rest of
the units than the required to spike, which translates into a weakening of the
synapse.  Conversely, if $L^i > 0$ some spontaneous evolution was required for
the unit $i$ to fire, Eq.~\eqref{eq:gradient} is positive and the synapse is
strengthened.  The intermediate case ($L^i = 0$), corresponds to $\eta=1$ and
no synaptic update is needed (nor is it defined). We will consider it thus $0$
for practical purposes.
\begin{figure}[!t]
\begin{center}
\includegraphics[width=.6\textwidth]{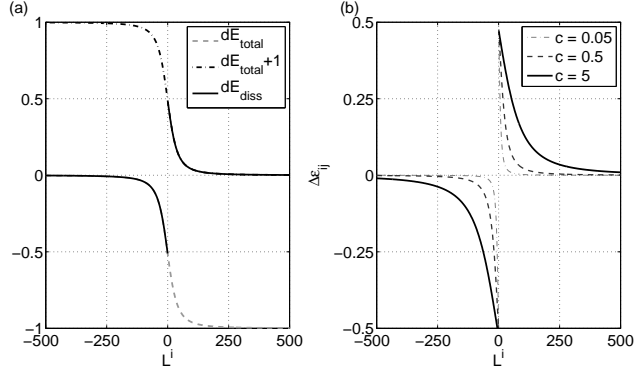} 
% This figure has been created using Tesis/code/neural/SimFreeSpont/figure_learning_rule.m script
% to reproduce it go to the figure_learning_rule dir
\end{center}
\caption[Derivative of $E_{diss}$ with respect to the global
coupling.]  { Plasticity rule. \textbf{(a)} First derivative of the
  dissipated spontaneous evolution $E_{diss}$ for $\kappa~=~1, L=1000$
  and $c=0.9$. \textbf{(b)} The same rule for different values of
  $c$.}
\label{fig:rule}
\end{figure}

%AK: I think her was something mixed up (eta>1 and eta>1):
Figure~\ref{fig:rule}a shows Eq.~\eqref{eq:gradient} in bold lines together
with $\partial{E^i_{total}}/\partial{\epsilon_{ij}}$ (dashed line,
corresponding to $\eta<1$) and
$\partial{E^i_{total}}/\partial{\epsilon_{ij}}+1$ (dashed-dotted, $\eta>1$),
for different values of the effective threshold $L^i$ of a given unit at the
	end on an ISI.  $E_{total}$ indicates the amount of synaptic change and
$E_{eff}$ determines whether the synapse is strengthened or weakened.
% As can be seen, the term corresponding to the total spontaneous
% evolution indicates the amount of change in the synapses and the
% term corresponding to the effective spontaneous evolution indicates
% whether the synapses must be strengthened or weakened.
The largest updates occur in the transition from a positive to a
negative $L^i$ and tend to zero for larger absolute values of $L^i$.
%This implies that significant updates correspond to those synapses
Therefore, significant updates correspond to those synapses with post-synaptic
neurons which during the last ISI have received a similar amount of activity
from the whole network as the one required to fire.
%Consequently, a network in a state far away from the critical
%point will require more steps to reach it than a network near the
%critical point.

We remark the similarity between Figure~\ref{fig:rule}b and the rule characterizing spike time dependent plasticity (STDP)~\cite{BiP98,SongMA00}.
%We notice a similarity in the shapes between the obtained function and those
%used in spike time dependent plasticity (STDP)~\cite{BiP98,SongMA00}.  i
Although in STDP the change in the synaptic conductances is determined by the
relative spike timing of the pre-synaptic neuron and the post-synaptic neuron
and here it is determined by $L^i$ at the spiking time of the post-synaptic
unit $i$, the largest changes in STDP occur also in an abrupt transition from
strengthening to weakening corresponding to $L^i=0$ in Figure~\ref{fig:rule}a.

%We notice a similarity in the shapes between the obtained function and
%those used in spike time dependent plasticity (STDP)~\cite{BiP98,SongMA00}.
%However,
%while in STDP the change in the synaptic conductances is determined by the
%relative spike timing of the pre-synaptic neuron and the post-synaptic neuron,
%here it is determined by $L^i$ at the spiking time of the post-synaptic unit
%$i$.
% If we replace the domains we get very similar functions.  On one
% hand, STDP long-term strengthening of synapses occurs if pre-synaptic
% action potentials precede post-synaptic firing by no more than about
% few milliseconds.  The analog situation in our model corresponds to
% the case where $L^i>0$ and the post-synaptic unit has not received
% enough spikes to reach the threshold in the last period.  On the
% other hand, pre-synaptic action potentials that follow post-synaptic
% spikes produce long-term weakening of synapses, which in our case
% corresponds to the situation where $L^i<0$ and the unit receives an
% excess of activity from the units in the network in the last period.
%Similar to our case,
% In
%STDP, however, the transition is produced when the time differences
%between pre- and post-synaptic action potentials passes through zero.

Figure~\ref{fig:rule}b illustrates the role of $c$ in the plasticity rule. For
small $c$, updates are only significant in a tiny range of $L^i$ values near
zero.  For higher values of $c$, the interval of relevant updates is widened.
The shape of the rule, however, is preserved, and the role of $c$ is just to
scale the change in the synapse. For the rest of this manuscript, we will use
$c = 1$.

\subsection{Simulations}
In this section we show empirical results for the
proposed plasticity rule.  We focus our analysis on the time
$\tau_{conv}$ required for the system to converge toward the critical
point.  In particular, we analyze how $\tau_{conv}$ depends on the
starting initial configuration and on the constant $\kappa$.
% For simplicity, we consider a global $\kappa$, but note that this is
% not a restriction, since different $\kappa_{ij}$ can be applied to
% each particular synapse $\epsilon_{ij}$.

For the experiments we use a network composed of $N=500$ units with
homogeneous $L=500$ and $p=0.9$.  Synapses are initialized
homogeneously and random initial states are chosen for all units
in each trial.  Every time a unit $i$ fires, we update its afferent
synapses $\epsilon_{ij}$, for all $j\neq i$, which breaks the
%AK: This fact breaks the
homogeneity in the interaction strengths.
% This figure has been created from Tesis/code/sim/plot_convergence_NIPS
% to reproduce it go to the figure_convergence dir
\begin{figure}[!t]
\begin{center}
  \includegraphics[width=.5\textwidth,angle=-90]{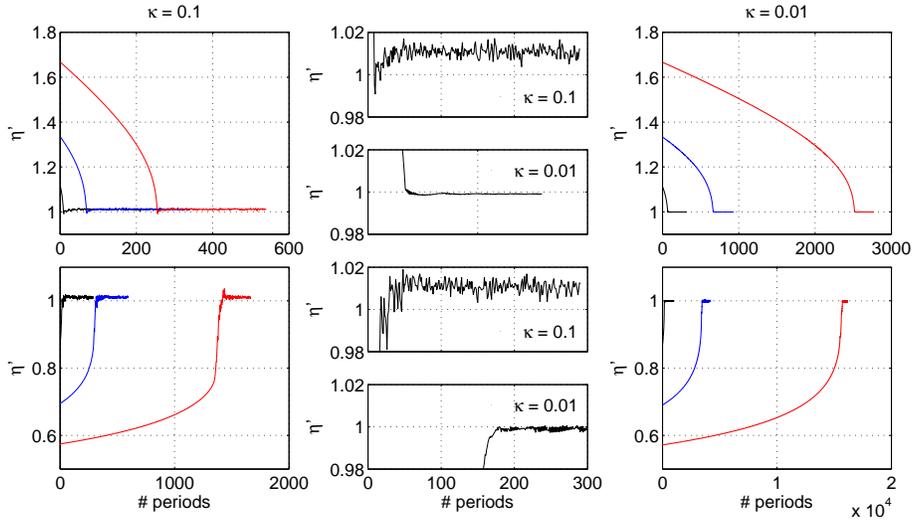}
\end{center}
\caption[Different realizations using the local learning rule for
$\eta>1$ and $\eta<1$.] 
{Empirical results of convergence toward $\eta=1$ for
  three different initial states above (top four plots) and below
  (bottom four plots) the critical point. Horizontal axis denote
  number of ISIs of the same random unit during all the simulations.
  On the left, results using the constant $\kappa=0.1$.  Larger panels
  shows the full trajectory until $10^3$ timesteps after convergence.
  Smaller panels are a zoom of the first trajectory $\eta_0 = 1.1$
  (top) and $\eta=0.87$ (bottom).  Right panels show the same type of
  results but using a smaller constant $\kappa = 0.01$.  }
\label{fig:above}
\end{figure}
The network starts with a certain initial condition $\eta_0$ and evolves
according to its original discrete dynamics, Eq.~\eqref{eq:discrete-final},
together with plasticity rule~\eqref{eq:rule}.  To measure the time
$\tau_{conv}$ necessary to reach a value close to $\eta=1$ for the first time,
we select a neuron~$i$ randomly and compute $\eta$ every time $i$ fires.  We
assume convergence when $\eta\in(1-\nu, 1+\nu)$ for the first time.  In these
initial experiments, $\nu$ is set to $\kappa/5$ and $\kappa$ is either $0.1$ or
$0.01$.

%We analyze $\tau_{conv}$ for a network initialized in the
%noise-dominated regime.
We performed $50$ random experiments for different initial configurations.  In
all cases, after an initial transient, the network settles close to $\eta=1$,
presenting some fluctuations.  These fluctuations did not grow even after
$10^6$ ISIs in all realizations.
%The probability distribution of the synaptic
%efficacies after convergence roughly follows a Gaussian distribution.
Figure~\ref{fig:above} shows examples for $\eta_0 \in \{0.58,0.7,0.87, 1.1,
1.3, 1.7\}$.

% Only data of the first $1000$ timesteps after convergence are shown.
%1000 ISIs or TIME-steps

%rate_eta_0.90_0.01.out
%rate_eta_0.75_0.01.out
%rate_eta_0.60_0.01.out
%rate_eta_1.75_0.01.out
%rate_eta_1.45_0.01.out
%rate_eta_1.15_0.01.out
We can see that for larger updates of the synapses ($\kappa=0.1$) the network
converges faster. However, fluctuations around the reached state, slightly
above $\eta=1$, are approximately one order of magnitude bigger than for
$\kappa=0.01$.
%Notice also that the reached state is slightly above $\eta=1$.
%This systematic shift is proportional to $\kappa$.
%
We therefore can conclude that $\kappa$ determines the speed of
convergence and the quality and stability of the dynamics at the critical
state: high values of $\kappa$ cause fast convergence but turn the dynamics of the
network less stable at the critical state.

We study now how $\tau_{conv}$ depends on $\eta_0$ in more detail.
% we use a simple analytical expression.  for the number of ISIs (or
% timesteps) required for convergence in function of an arbitrary
% choice for $\eta_0$.
Given $N, L, c$ and $\kappa$, we can approximate the global change in $\eta$
after one entire ISI of a random unit assuming that all neurons change its
afferent synapses uniformly. This gives us a recursive definition for the
sequence of $\eta_t$s generated by the synaptic plasticity rule:
% \begin{align}\label{eq:eta_rec}
%   \Delta(\eta_t) & = \kappa(N-1)\left( \frac{
%       -\frac{L_{eff}(\eta_t)}{2c} - \frac{1}{4} } { \sqrt{ \left(
%           \frac{L_{eff}(\eta_t)}{c} + 1\right)^2 +
%         \frac{L-L_{eff}(\eta_t)}{c}}} - \frac{1}{2} +\begin{cases} 0 &
%       \text{if $\eta_t < 1$} \\ 1 & \text{if $\eta_t > 1$}
%   \end{cases}\right),\\
% & \text{where } L_{eff}(\eta_t) = (L-1)\left(1-\frac{1}{\eta_t}\right)
% \quad \text{ and } \quad \eta_{t+1} = \eta_t+\Delta(\eta_t) .
% \end{align}
\begin{align*}
  \Delta(\eta_t) & = \kappa(N-1)\left( \frac{
      -L_{eff}(\eta_t) - c } { 2 \sqrt{ \left(
          L_{eff}(\eta_t) + 2c\right)^2 +
        2c(L-L_{eff}(\eta_t))}} + \frac{\text{sgn}(\eta_t-1)}{2}\right),
\end{align*}				
\begin{align*}
& \text{where } L_{eff}(\eta_t) = (L-1)\left(1-\frac{1}{\eta_t}\right)
\quad \text{ and } \quad \eta_{t+1} = \eta_t+\Delta(\eta_t) .\notag
\end{align*}
\begin{figure}[!t]
\begin{center}
  \includegraphics[width=.65\textwidth]{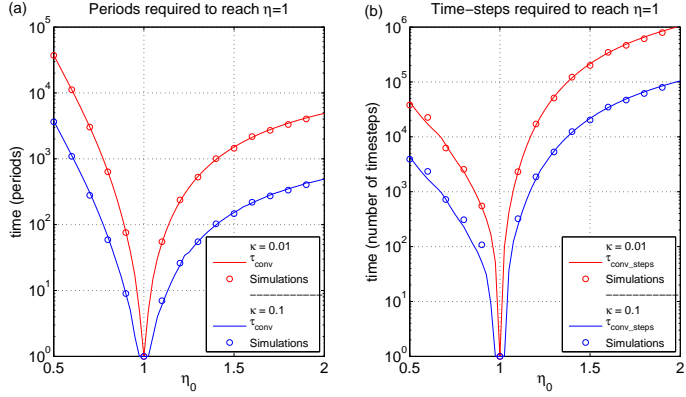}
\end{center}
\caption[Time until the critical point is reached depending on the 
initial $\eta_0$.]  { Number of ISIs \textbf{(a)} and timesteps
  \textbf{(b)} required to reach the critical state in 
  function of the initial configuration $\eta_0$.  Rounded dots 
  indicate empirical results as averages over $10$ different 
  realizations starting from the same $\eta_0$.  Continuous curves
  correspond to Eq.~\eqref{eq:periods}.
  Parameter values are $N=500, L=500, p=0.9, c=1$, $\nu=\kappa/5$.
}
\label{fig:all_etas}
\end{figure}
% The number of periods can then be obtained by means of the following recursive equation:
% \begin{align}\label{eq:periods}
% \tau_{conv}(\eta_{t}) & = \begin{cases}
% 1                                   & \text{if $|\eta_{t}-1| \leq \nu$}\\
% 1 + \tau_{conv}( \eta_{t+1})  & \text{otherwise.}
% \end{cases}
% \end{align}
%Then the number of ISIs is given by
%\begin{align}\label{eq:periods}
%\tau_{conv}&=\text{min}(\{i:|\eta_{t}-1| \leq \nu\})
%\end{align}
%% and the number of timesteps can be obtained by the following
%% recursive equation which uses the lower bound approximation of the
%% mean period $\tau$:
%% \begin{align}\label{eq:steps}
%% \tau_{conv\_steps}(\eta_{t}) & = \begin{cases}
%% 0             & \text{if $|\eta_{t}-1| \leq \nu$}\\
%% \langle\tau\rangle_{min}(\eta_{t}) + \tau_{conv\_steps}( \eta_{t+1})  & \text{otherwise.}
%% \end{cases}
%% \end{align}
%and the number of timesteps can be obtained from the approximation
%$\langle\tau\rangle_{min}$ of the mean ISI $\langle\tau\rangle$:
%\begin{align}\label{eq:steps}
%\tau_{conv\_steps} & = \sum_{t=0}^{\tau_{conv}} \langle\tau\rangle_{min}(\eta_{t}).
%\end{align}
Then the number of ISIs and the number of timesteps can be obtained by\footnote{The value of $\tau_{app}(\eta_{t})$ has to be calculated using an
$\langle\epsilon\rangle$ corresponding to $\eta_{t}$ in
Eq.~\eqref{con:tau_approx}.}:
\begin{align}\label{eq:periods}
\tau_{conv}&=\text{min}(\{i:|\eta_{t}-1| \leq \nu\}),
& \tau_{conv\_steps} & = \sum_{t=0}^{\tau_{conv}} \tau_{app}(\eta_{t}).
\end{align}
%
%
%Then the number of ISIs and the number of timesteps can be obtained by:
%\begin{align}\label{eq:periods}
%\tau_{conv}&=\text{min}(\{i:|\eta_{t}-1| \leq \nu\}),
%& \tau_{conv\_steps} & = \sum_{t=0}^{\tau_{conv}} \langle\tau\rangle_{min}(\eta_{t}).
%\end{align}
Figure~\ref{fig:all_etas} shows empirical values of $\tau_{conv}$ and
$\tau_{conv\_steps}$ for several values of $\eta_0$ together with the
approximations~\eqref{eq:periods}. Despite the inhomogeneous coupling
strengths, the analytical approximations (continuous lines) of the experiments
(circles) are quite accurate.  Typically, for $\eta_0<1$ more spikes are
required for convergence than for $\eta_0>1$. However, the opposite occurs if
we consider timesteps as time units.  A hysteresis effect (described in
\cite{Kaltenbrunner2007NECO}) present in the system if $\eta_0 < 1$, causes the
system to be more resistant against synaptic changes, which increases the
number of updates (spikes) necessary to achieve the same effect as for $\eta_0
> 1$.  Nevertheless, since the ISIs are much shorter for supercritical coupling
the actual number of time steps is still lower than for subcritical coupling.

\section{Discussion}
\label{sec:neural-concl}
Based on the amount of spontaneous evolution which is dissipated during an ISI,
we have derived a local synaptic mechanism which causes a network of spiking
neurons to self-organize near a critical state. Our motivation differs from
those of similar studies, for instance~\cite{Kinouch2006}, where the average
branching ratio $\sigma$ of the network is used to characterize criticality.
Briefly, $\sigma$ is defined as the average number of excitations created in
the next time step by a spike of a given neuron.

The inverse of $\eta$ plays the role of the branching ratio $\sigma$ in our
model.  If we initialize the units uniformly in $[1,L]$, we have approximately
one unit in every subinterval of length $\eta\epsilon$, and in consequence, the
closest unit to the threshold spikes in $1/\eta$ cases if it receives a spike.
% need $1/\eta$ spikes to surpass the threshold for the closest unit on average.
For $\eta > 1$, a spike of a neuron rarely induces another neuron to spike, so
$\sigma < 1$.  Conversely, for $\eta < 1$, the spike of a single neuron
triggers more than one neuron to spike ($\sigma > 1$). Only for $\eta = 1$ the
spike of a neuron elicits the order of one spike ($\sigma=1$). Our study thus
represents a realization of a local synaptic mechanism which induces global
homeostasis towards an optimal branching factor.

This idea is also related to the SOC rule proposed in \cite{Bertschinger},
where a mechanism is defined for threshold gates (binary units) in terms of bit
flip probabilities instead of spiking neurons.  As in our model, criticality is
achieved via synaptic scaling, where each neuron adjusts its synaptic input
according to an effective threshold called \emph{margin}.% in \cite{Bertschinger}.
%where self-organization is achieved via synaptic scaling
%The rule derived here for the case of spiking neurons is
%therefore based in the same mechanism.

%Although formulated
%for circuits consisting of threshold gates (no memory) in terms of bit flip
%probabilities, the basic mechanism of adjusting the input according to an
%effective threshold, in \cite{Bertschinger} called \emph{margin}, is very
%similar.
%Criticality in both models is achieved
%if the spiking / bit flip probability approaches 1 during the last time step
%	(because of the lack of an integrator) / the last ISI.

%The inverse of $\eta$ seems to play the role of $\sigma$ in our model if we
%consider the ratio between the number of spikes triggered by a neuron and the
%mean ISI of the network.  For $\eta > 1$, a spike of a neuron rarely
%induces another neuron to spike,
%%ratio between number of neurons N and the ISI (is large compared with),
%so $\sigma < 1$.  Conversely, for $\eta < 1$, the spike of a single neuron
%triggers more activity than the mean ISI, thus $\sigma > 1$. Only for $\eta =
%1$ the spike of a neuron elicits the order of $\sqrt{N}$ spikes, which
%coincides with the mean ISI~\cite{Kaltenbrunner2007NECO}, resulting in a
%branching ratio $\sigma=1$. Our study thus represents a realization of a local
%synaptic mechanism which induces global homeostasis towards an optimal
%branching factor.

%AK: comma added
When the network is operating at the critical regime, the dynamics can be seen
as balancing between a predictable pattern of activity and uncorrelated random
behavior typically present in SOC.  One would also expect to find
macroscopic magnitudes distributed according to scale-free distributions.
Preliminary results indicate that, if the stochastic evolution is reset to zero
($p=0$) at the critical state, inducing an artificial spike on a randomly
selected unit causes neuronal avalanches of sizes and lengths which span
several orders of magnitude and follow heavy tailed distributions.  These
results are in concordance with what is usually found for SOC and
will be published elsewhere.

The spontaneous evolution can be interpreted for instance as activity from
other brain areas not considered in the pool of the simulated units, or as
stochastic sensory input. Our results indicate that the amount of this
stochastic activity that is absorbed by the system is maximized at an optimal
state, which in a sense minimizes the possible effect of fluctuations due to
noise on the behavior of the system.

The application of the synaptic rule for information processing is
left for future research. We advance, however, that external perturbations 
when the network is critical would cause a transient activity.  During
the transient, synapses could be modified according to some other form of
learning to encode the proper values which drive the whole network to attain a
characteristic synchronized pattern for the external stimuli presented.  We
conjecture that the hysteresis effect shown in the regime of $\eta<1$ may be
suitable for such purposes, since the network then is able to keep the same
pattern of activity until the critical state is reached again.

\subsubsection*{Acknowledgments}We thank Joaqu\'in J. Torres and Max Welling for
useful suggestions and interesting discussions.
%We thank anonymous reviewers for his suggestions which helped to improve this work.
%This research is part of the Interactive Collaborative In-
%formation Systems (ICIS) project, supported by the Dutch
%Ministry of Economic Affairs, grant BSIK03024.
%%Use unnumbered third level headings for the acknowledgments. All
%%acknowledgments go at the end of the paper.

%\subsubsection*{References}

{\small
\bibliographystyle{plain}
\bibliography{gomez-nips08}
}
%\end{thebibliography}

\end{document}